\documentclass[journal,twoside,web]{ieeecolor}
\usepackage{tmi}
\usepackage{cite}
\usepackage{amsmath,amssymb,amsfonts}
\usepackage{algorithmic}
\usepackage{graphicx}
\usepackage{textcomp}

\usepackage{booktabs}
\usepackage{multirow}
\usepackage{array}
\newcolumntype{P}[1]{>{\centering\arraybackslash}p{#1}}
\newcolumntype{M}[1]{>{\centering\arraybackslash}m{#1}}

\usepackage{pifont}
\usepackage{bbding}

\usepackage{hyperref}

\def\BibTeX{{\rm B\kern-.05em{\sc i\kern-.025em b}\kern-.08em
    T\kern-.1667em\lower.7ex\hbox{E}\kern-.125emX}}
\begin{document}
\title{Unlocking the diagnostic potential of electrocardiograms through information transfer from cardiac magnetic resonance imaging}

\author{Özgün Turgut, Philip Müller, Paul Hager, Suprosanna Shit, Sophie Starck, \\Martin J. Menten, Eimo Martens, and Daniel Rueckert
\thanks{Özgün Turgut, Philip Müller, Paul Hager, Sophie Starck, Martin J. Menten, and Daniel Rueckert are with the School of Computation, Information and Technology, Technical University of Munich, Germany (e-mail: oezguen.turgut@tum.de, philip.j.mueller@tum.de, paul.hager@tum.de, sophie.starck@tum.de, martin.menten@tum.de, daniel.rueckert@tum.de). 
Suprosanna Shit is with the Department of Quantitative Biomedicine, University of Zurich, Switzerland (suprosanna.shit@uzh.ch).
Özgün Turgut, Philip Müller, Paul Hager, Sophie Starck, Eimo Martens, and Daniel Rueckert are with the School of Medicine, Klinikum Rechts der Isar, Technical University of Munich, Germany (e-mail: eimo.martens@mri.tum.de). 
Martin J. Menten and Daniel Rueckert are also with the Department of Computing, Imperial College London, United Kingdom, and the Munich Center for Machine Learning, Munich, Germany. 
This research has been conducted using the UK Biobank Resource under Application Number 87802.}}

\maketitle
\begin{abstract}
Cardiovascular diseases (CVD) can be diagnosed using various diagnostic modalities. 
The electrocardiogram (ECG) is a cost-effective and widely available diagnostic aid that provides functional information of the heart. 
However, its ability to classify and spatially localise CVD is limited. 
In contrast, cardiac magnetic resonance (CMR) imaging provides detailed structural information of the heart and thus enables evidence-based diagnosis of CVD, but long scan times and high costs limit its use in clinical routine. 
In this work, we present a deep learning strategy for cost-effective and comprehensive cardiac screening solely from ECG. 
Our approach combines multimodal contrastive learning with masked data modelling to transfer domain-specific information from CMR imaging to ECG representations. 
In extensive experiments using data from 40,044 UK Biobank subjects, we demonstrate the utility and generalisability of our method for subject-specific risk prediction of CVD and the prediction of cardiac phenotypes using only ECG data. 
Specifically, our novel multimodal pre-training paradigm improves performance by up to 12.19$\,$\% for risk prediction and 27.59$\,$\% for phenotype prediction.
In a qualitative analysis, we demonstrate that our learned ECG representations incorporate information from CMR image regions of interest. 
Our entire pipeline is publicly available at \url{https://github.com/oetu/MMCL-ECG-CMR}.
\end{abstract}


\section{Introduction}
\label{sec:introduction}
The standard $12$-lead electrocardiogram (ECG) is a widely available clinical modality due to its ease of use, short acquisition time, and cost-effectiveness. 
It is a non-invasive method used by cardiologists to record the heart's electrical activity. 
In clinical routine, the ECG serves as a valuable diagnostic aid, as it can reveal characteristic signatures associated with cardiac disorders, including atrial fibrillation and left ventricular systolic dysfunction \cite{Siontis2021}. 
However, a detailed assessment of morphological conditions from ECG requires the inverse modelling of the heart's electrical activity to the actual physiological source, which is known to be an ill-posed problem \cite{Gulrajani1998, Schenone2016}.
The characterisation and spatial localisation of cardiac disorders is thus limited to well investigated conditions, such as myocardial infarction, which can be localised even using simple statistical analysis \cite{Engelen1999, Xiong2021}.

\begin{figure*}[ht!]
    \centering
    \includegraphics[width=0.99\textwidth]{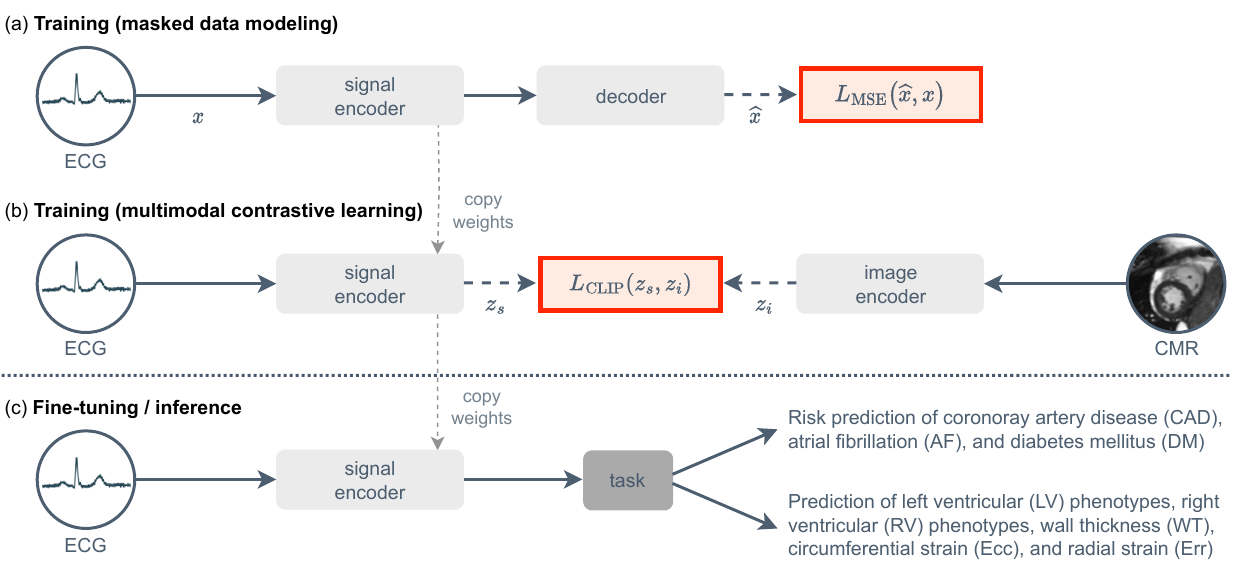}
    \caption{Overview of our training stages and inference process. (a) Our proposed approach uses masked data modelling to learn meaningful ECG representations, eliminating redundancy inherent to standard $12$-lead ECG. (b) We introduce multimodal contrastive learning to transfer domain-specific information from CMR imaging to ECG. (c) Once pre-trained, the signal encoder is fine-tuned and can be used during inference to predict the risk of cardiovascular diseases and to predict cardiac phenotypes solely from ECG.}
    \label{fig:overview}
\end{figure*}
\textcolor{black}{In contrast,} cardiac magnetic resonance (CMR) imaging is a modality that provides high-resolution, volumetric images for a precise assessment of the cardiac morphology. It allows for accurate prediction of important cardiac phenotypes, such as left ventricular ejection fraction \textcolor{black}{(LVEF)} or left ventricular volumes\textcolor{black}{, including end-diastolic (LVEDV), end-systolic (LVESV), and stroke volume (LVSV),} that can be used to quantitatively assess heart failure \cite{Pombo1971} \textcolor{black}{\cite{Avendi2016, Duffy2022}}. As a result, CMR imaging serves as the gold-standard for evidence-based diagnosis of various cardiac disorders \cite{Lee2018}, 
\textcolor{black}{enabling detailed assessment of conditions such as coronary artery disease (CAD) \cite{Reynolds2021, Wang2024}, atrial fibrillation (AF) \cite{Oakes2009, Bertelsen2020}, or diabetes mellitus (DM) \cite{Emerging2010, Sorensen2020, Ng2021}.}
However, its application in clinical practice is limited due to long scan times, high associated costs, and the necessity for trained personnel \cite{Von2017}. 

In this study, we present a deep learning strategy that combines the availability of ECG with the informative value of CMR imaging to provide a cost-effective and holistic cardiac screening tool for clinical routine. We introduce a new contrastive learning strategy that processes large amounts of unlabeled, multimodal biobank data in a self-supervised fashion. Initially, we pre-train unimodal encoders on paired ECG and CMR imaging data to transfer complementary information across modalities. Afterwards, the trained signal encoder is used to predict various cardiac conditions and phenotypes using ECG data only. A graphical visualisation of our approach is given in Figure \ref{fig:overview}. 
Through information transfer from CMR imaging to ECG, we reduce the \textcolor{black}{reliance} on costly CMR \textcolor{black}{scans} during inference and unlock the diagnostic potential of ECG for affordable care for patients with cardiovascular conditions. 
To this end, our key contributions are as follows:

\begin{enumerate}
    \item We propose \textcolor{black}{a novel multimodal pre-training paradigm} that utilises $12$-lead ECG and CMR imaging. We combine multimodal contrastive learning with masked data modelling to transfer information from CMR imaging to ECG.
    \item In extensive benchmarking experiments using $40,044$ subjects from the UK Biobank \cite{Sudlow2015}, we show that our learned ECG representations generalise across various downstream applications. After fine-tuning, they can be used to predict risks of various cardiovascular diseases and to determine distinct cardiac phenotypes. Our method outperforms state-of-the-art self-supervised and supervised baselines, which use either ECG, CMR imaging, or tabular data for their predictions.
    \item We quantitatively demonstrate the outsized importance of masked data modelling in the information transfer process with extensive ablation studies. Further, we qualitatively show the information transfer from CMR imaging to ECG using latent vector similarity measures.
    \item We publish our entire pipeline as an open source tool, including the pre-trained model weights at \url{https://github.com/oetu/MMCL-ECG-CMR}.
\end{enumerate}

\section{Related works}
\label{sec:related_works}
\subsection{Deep-learning-based processing of ECG} 
The application of deep learning to ECG is increasingly becoming popular in cardiovascular medicine \cite{Siontis2021}. However, most of the previous works investigate supervised approaches to improve classification performance \cite{Attia2019, Raghunath2020, Zhu2020}, which rely on expensive annotation by medical experts. Beyond supervised learning, self-supervised methods \cite{Sarkar2020, Zhang2022} have been proposed to eliminate this data dependency and learn ECG representations using unlabeled data. Implementations such as CLOCS \cite{Kiyasseh2021} have been introduced to tailor contrastive learning to medical time series data, enabling self-supervised representation learning of standard $12$-lead ECG. 
\textcolor{black}{Nevertheless, most previous works study the utility of task-specific ECG representations for prediction of either specific cardiac disorders, such as AF \cite{Khurshid2022, Yuan2023} or CAD \cite{Jahmunah2021, Awasthi2023}, or specific cardiac imaging phenotypes, such as left ventricular mass (LVM) \cite{Khurshid2021} or LVEF \cite{Vaid2023}.
In contrast, our work aims to explore the potential of \textit{general-purpose} ECG representations that translate to a broad range of tasks, including the risk prediction of various cardiac diseases and the prediction of distinct cardiac imaging phenotypes.}

\subsection{Masked data modelling}
\begin{figure*}[ht!]
    \centering
    \includegraphics[width=0.99\textwidth]{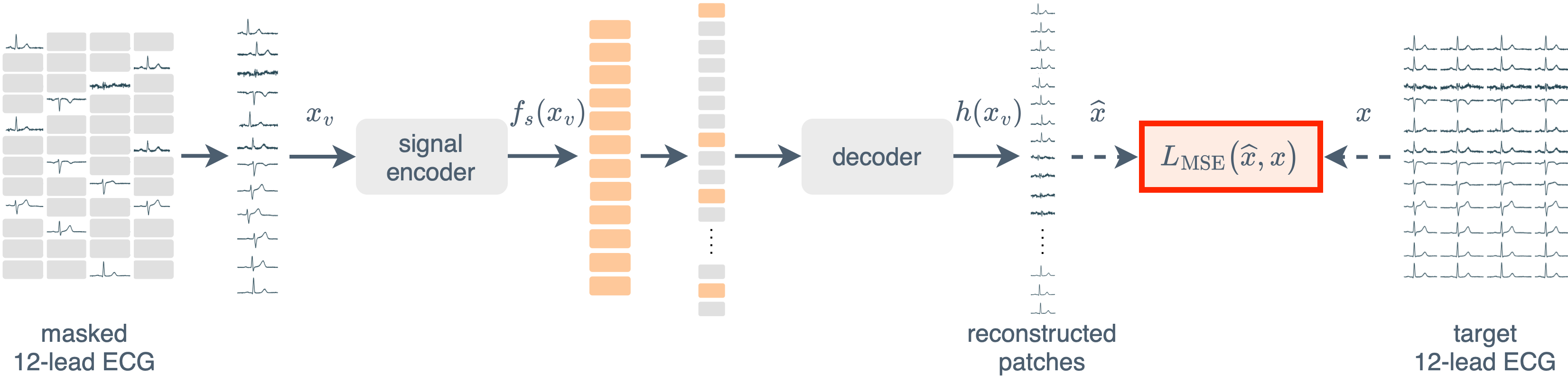}
    \caption{We use masked data modelling to eliminate redundancy inherent to 12-lead ECG, thus generating meaningful ECG representations. To this end, we split the ECG data into patches of predefined size, out of which a random set is masked out. 
    \textcolor{black}{Note that for visualisation purposes the patch size is set to cover a single heartbeat, however, the actual patch size may vary.}
    Only the small subset of visible patches is encoded by the signal encoder. The full set of encoded and masked patches is reconstructed by the decoder.
    }
    \label{fig:mae}
\end{figure*}
Masked data modelling approaches have proven effective for self-supervised representation learning. \textcolor{black}{With the goal of recovering} a random set of masked data points from the visible data, these self-supervised approaches can be referred to as masked autoencoding \cite{Chen2020Generative, Pathak2016, Vincent2010} techniques. Initially introduced for self-supervised pre-training in natural language processing (NLP) \cite{Devlin2018}, masked data modelling has \textcolor{black}{since} been extended to other modalities such as imaging \cite{He2022}, video \cite{Feichtenhofer2022}, and audio \cite{Huang2022}. Unlike contrastive learning \cite{Chen2020} and most other self-supervised pre-training strategies, masked data modelling does not rely on heavy data augmentations, which are challenging to design for sequential data \cite{Assran2023}. Furthermore, it enables learning representations that generalise well to diverse downstream applications \cite{Brown2020}. 

Recent works investigating masked data modelling, such as iGPT \cite{Chen2020Generative}, ViT \cite{Dosovitskiy2020}, BEiT \cite{Bao2022}, or the masked autoencoder (MAE) \cite{He2022}, are relying on Transformers \cite{Vaswani2017} as backbone architecture and thus follow the example of BERT \cite{Devlin2018} in NLP. Transformers have shown great success in modelling local and global dependencies, enabling the model to capture relationships even in long sequential data \cite{Beltagy2020}. In the medical domain, \cite{Zhang2022} have recently applied the MAE to detect arrhythmia from ECG. We also employ a MAE during the initial step of our approach and further introduce unstructured masking similar to \cite{Huang2022} to learn meaningful ECG representations.

\subsection{Multimodal contrastive learning} 
Contrastive learning frameworks such as SimCLR \cite{Chen2020}, BYOL \cite{Grill2020}, or BarlowTwins \cite{Zbontar2021} are able to learn rich representations of imaging data. Leveraging a contrastive loss \cite{Hadsell2006}, these approaches align augmented views of a sample in the latent space. Motivated by the success in learning unimodal representations, contrastive learning has been recently extended to consider data from multiple modalities. Implementations such as CLIP \cite{Radford2021} and ConVIRT \cite{Zhang2022conVIRT} align representations of paired imaging and text data. Other approaches apply contrastive techniques to learn general-purpose representations using two different imaging modalities \cite{Pielawski2020}, imaging and tabular data \cite{Hager2023}, video and text \cite{Xu2021, Zolfaghari2021}, and video and audio \cite{Wang2021, Ma2021}. 
\textcolor{black}{Recent work by \cite{Radhakrishnan2023}}
\textcolor{black}{introduced}
\textcolor{black}{a multimodal contrastive learning approach that incorporates medical time series data,} 
\textcolor{black}{i.e.}
\textcolor{black}{$12$-lead ECG, and imaging data,}
\textcolor{black}{i.e.}
\textcolor{black}{CMR imaging. However,} 
\textcolor{black}{in certain low data scenarios,} 
\textcolor{black}{their approach}
\textcolor{black}{still underperforms compared} 
\textcolor{black}{to ECG-only solutions, which do not}
\textcolor{black}{rely on}
\textcolor{black}{expensive CMR imaging data.}
This highlights the clear need for strategies that effectively transfer morphological information from CMR imaging to ECG representations, which we aim to address in this work.
 
\section{Methods}
\begin{figure*}[ht!]
    \centering
    \includegraphics[width=0.99\textwidth]{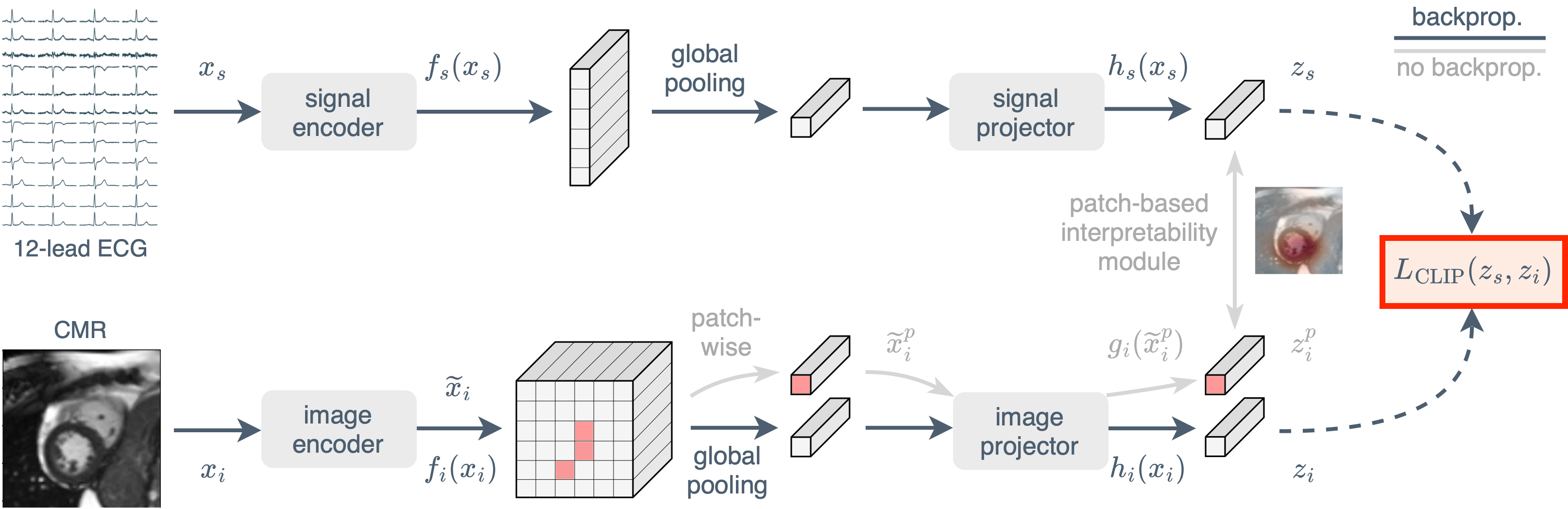}
    \caption{We introduce multimodal contrastive learning that combines 12-lead ECG and CMR imaging, enabling self-supervised information transfer from CMR imaging to ECG. ECG and CMR images are embedded separately using unimodal encoders. The representations are projected separately onto a shared latent space, where information is exchanged between both modalities. A passive interpretability module visualises the similarity between the global ECG representation and local CMR image representations, allowing for a qualitative evaluation of the information transfer.}
    \label{fig:clip}
\end{figure*}
In this work, we present a novel \textcolor{black}{multimodal pre-training paradigm} that incorporates time series and imaging data to train unimodal encoders (see Figure \ref{fig:overview}). We use masked data modelling (MDM) to train a signal encoder unimodally on large amounts of unlabeled $12$-lead ECG, allowing for learning meaningful ECG representations as described in Section \ref{sec:mdm}. After pre-training solely on ECG data, we introduce multi-modal contrastive learning (MMCL) to further pre-train our unimodal signal encoder with information from CMR imaging as described in Section \ref{sec:mmcl}. We introduce an interpretability module in Section \ref{sec:interpretability_module} to visually assess the information transfer from CMR imaging to ECG. After these pre-training steps, the unimodal signal encoder can be fine-tuned on a limited amount of labeled data to predict the subject-specific risk of cardiovascular diseases and to determine cardiac imaging phenotypes during inference solely from ECG.

\subsection{Masked data modelling}
\label{sec:mdm}
Given the redundancy inherent to standard $12$-lead ECG, we first pre-train our signal encoder unimodally on large amounts of unlabeled ECG data to learn meaningful representations. To this end, we use masked data modelling (MDM) with a high masking ratio $\rho$. This allows us to eliminate redundancy across signal channels and time without relying on heavy data augmentations, which are difficult to design for time series data \cite{Assran2023}. We choose the Transformer \cite{Vaswani2017} as our backbone architecture to capture local and global dependencies in multi-channel ECG. We base our MDM implementation on the MAE \cite{He2022}. A graphical visualisation of the method can be found in Figure \ref{fig:mae}.

Assume a multi-channel time series sample \textcolor{black}{$\bar{x} \in \mathbb{R}^{C \times T}$}, where $C$ and $T$ denote the number of signal channels and time points, respectively.
We split the time series into $T'$ temporal patches of size $P$ along the time dimension, resulting in $N = C \cdot T'$ patches ${x'}_{c,t} \in \mathbb{R}^{1\times P}$, with $c \in \{1, 2, ..., C\}$ and $t \in \{1, 2, ..., T'\}$.
Next, we use a 1D convolutional layer to project these patches into a latent space of dimension $D$, such that ${x''}_{c,t} \in \mathbb{R}^{1\times D}$.
We add standard 2D sinusoidal positional embeddings $e_{c,t} \in \mathbb{R}^{1\times D}$ \cite{Vaswani2017} to  adequately model the channel and time position, such that $x'''_{c,t} = x''_{c,t} + e_{c,t}$.
Eventually, we flatten all patches into a 1D sequence $x \in \mathbb{R}^{N \times D}$.

Assume a randomly drawn binary mask $m\in\{0,1\}^N$ ($0$ drawn with probability $\rho$) that is applied to $x$ to obtain a visible view $x_v=x[m]\in\mathbb{R}^{N_v \times D}$ and a complementary masked view $x_m=x[1-m]\in\mathbb{R}^{N_m \times D}$, where $N=N_v+N_m$. Assume two randomly drawn views $x_v=x[m], x'_v=x[m']$ of $x$, where $m \neq m'$. Let $f_s(\cdot)$ and $g(\cdot)$ be the signal encoder and decoder, respectively, such that the MAE can be represented as $h(\cdot) = (g \circ f_s)(\cdot)$, where 
\begin{equation}
\label{eq:x_hat}
\hat{x} = h(x_v) = g\left(f_s(x_v)\right).
\end{equation}

We train the MAE by optimising the mean squared error (MSE) loss, which can be lower bounded by a global alignment loss with respect to the representations $f_s(x_v)$ and $f_s(x'_v)$, under the assumption that $g(\cdot)$ is $L$-bi-Lipschitz \cite{Zhang2022Exp}: 
\begin{subequations}
\begin{align}
\label{eq:mse}
\mathcal{L}_\text{MSE} &= \mathbb{E}_{p(x)}\mathbb{E}_{p(x_v|x)} || g(f_s(x_v)) - x ||^2 \\
\label{eq:align}
&\ge - \frac{1}{2L} \mathbb{E}_{p(x_v,x'_v|x)} f_s(x_v)^\top f_s(x'_v) - \epsilon + \text{const}.
\end{align}
\end{subequations}
Minimising the MSE loss (Equation \ref{eq:mse}) enforces the encoder $f_s(\cdot)$ to focus on \textit{local} features of the visible view $x_v$ that are crucial for reconstructing the entire sample $x$, i.e., features that also generalise to the masked view $x_m$. At the same time, the global alignment loss (Equation \ref{eq:align}) enforces $f_s(\cdot)$ to consider \textit{global} features of the visible view $x_v$ that generalise to any other visible view $x'_v$ of $x$. Taken together, optimising the MSE loss eliminates redundancy and enables learning a signal encoder $f_s(\cdot)$ that creates representations with high information density.

\subsection{Multimodal contrastive learning}
\label{sec:mmcl}
We introduce multimodal contrastive learning (MMCL) incorporating both $12$-lead ECG and CMR imaging, to enable self-supervised information transfer from CMR imaging to ECG. 
We base our multimodal solution on SimCLR \cite{Chen2020}. 
A graphical visualisation of our approach is shown in Figure \ref{fig:clip}.

Assume pairs of time series $x_s$ and imaging data $x_i$, first augmented and then processed by a signal encoder $f_s(\cdot)$ and an image encoder $f_i(\cdot)$, respectively. 
The resulting representations $f_s(x_s)$ and $f_i(x_i)$ are passed through separate projectors $g_s(\cdot)$ and $g_i(\cdot)$ to generate the projections $z_s = g_s(f_s(x_s))$ and $z_i = g_i(f_i(x_i))$, respectively, which are $\ell_2$-normalised and mapped onto a shared latent space. $f_s(\cdot)$ and $g_s(\cdot)$ can be jointly represented as $h_s(\cdot) = (g_s \circ f_s)(\cdot)$, such that
\begin{equation}
\label{eq:proj_emb}
z_s = h_s(x_s) = g_s(f_s(x_s)).
\end{equation}
$h_i(\cdot)$ is defined analogously. In the shared latent space, we optimise the CLIP loss \cite{Radford2021} to maximise the similarity of paired time series data $x_s$ and imaging data $x_i$ in a batch $B$, while minimising the similarity of unpaired data. Unlike the original InfoNCE loss \cite{Oord2018}, the CLIP loss \cite{Radford2021} exclusively contrasts projections between different modalities and does not contrast data within one modality. The loss for the signal modality is defined as
\begin{equation}
\label{eq:l_sig}
\mathcal{L}^{\text{sig}} = \mathbb{E}_{p(x_s, x_i)}\mathbb{E}_{p(x^j_i)} \left[ - \log \frac{\exp\left(z_s^\top z_i / \tau\right)}{\sum_{j \in B} \exp\left(z_s^\top z^j_i / \tau\right)} \right],
\end{equation}
where $\tau$ denotes a temperature parameter, and the loss for the imaging modality as
\begin{equation}
\label{eq:l_img}
\mathcal{L}^{\text{img}} = \mathbb{E}_{p(x_i, x_s)}\mathbb{E}_{p(x^j_s)} \left[ - \log \frac{\exp\left(z_i^\top z_s / \tau\right)}{\sum_{j \in B} \exp\left(z_i^\top z^j_s / \tau\right)} \right].
\end{equation}
The total loss is thus
\begin{equation}
\mathcal{L}_\text{CLIP} = (1-\lambda)\mathcal{L}^{\text{sig}} + \lambda \mathcal{L}^{\text{img}},
\end{equation}
where $\lambda$ denotes a parameter to balance the loss between signal and imaging modality. 
\textcolor{black}{Note that $\lambda$ is introduced to potentially shift the supervision during multimodal pre-training towards the modality with higher information density.}

\subsection{Patch-based interpretability module}
\label{sec:interpretability_module}
To qualitatively evaluate the information transfer during MMCL, we introduce a patch-based interpretability module similar to \cite{Huang2021, Boecking2022}. 
\textcolor{black}{While these works apply local losses or even require ground truth bounding boxes, our module is non-trainable to provide zero-shot interpretation of the information transfer from CMR imaging to ECG without the need for additional resources.}
Specifically, our passive module 
\textcolor{black}{determines the cosine similarity between the global ECG representation and local CMR region representations, generating a similarity grid that is overlaid} 
onto the CMR image as heatmap. Assume the embedded image $\widetilde{x}_i = f_i(x_i) \in \mathbb{R}^{H\times W\times D_i}$, where $H, W, D_i$ are the height, width, and dimension of the embedding, respectively. We project each local image patch $\widetilde{x}^p_i \in \mathbb{R}^{1\times1\times D_i}$, where $p \in \{(h, w)|1\leq h\leq H \land 1\leq w\leq W\}$, separately to the latent space. In the shared latent space, we then determine the cosine similarity between each local image projection $z^p_i = g_i(\widetilde{x}^p_i)$ and the signal projection $z_s = h_s(x_s)$. The resulting similarity grid offers insights into CMR image regions that correlate with the ECG.

\section{Experiments}
\subsection{Dataset and pre-processing}
We train our multimodal method using pairs of $12$-lead ECG and CMR imaging data from the UK Biobank population study \cite{Sudlow2015}. The ECG and CMR images were acquired within a few hours during the subject's visit, thus capturing similar diagnostic information about the cardiovascular health. We use routine $10$-s, $12$-lead ECG 
\textcolor{black}{recordings sampled at a} 
frequency of $500$\,Hz. The baseline drift was removed using the asymmetric least square smoothing technique \cite{Zhang2010}. We apply standard normalisation separately to the Einthoven, Goldberger, and Wilson leads \cite{Khan2008}. 

We use short-axis CMR images that provide a cross-sectional view of the heart's left and right ventricle. Short-axis images were selected as they depict the left ventricle, in which early warning signs of cardiac dysfunction may become evident \cite{Raisi2021, Tsao2015}. The imaging data contains three-channel two-dimensional images, whose channels are the middle baso-apical slice at the end-systolic phase, the end-diastolic phase, and the mid-phase. The images were zero-padded to $210\times210$ pixels and min-max-normalised between 0 and 1. We manually cropped the images around the visible heart and resized them back to the original size.

The \textcolor{black}{full} dataset contains $40,044$ unique subjects with pairs of ECG and CMR imaging data, which we split into $28,030$ training, $6,007$ validation, and $6,007$ testing pairs.

\subsection{Prediction targets}
\label{sec:prediction_targets}
Our aim is to predict the subject-specific risk of various cardiovascular diseases defined by the International Classification of Diseases (ICD$10$) code solely using ECG data. In this study, we investigate coronary artery disease (CAD) (ICD$10$ codes I$20$-I$25$), which is closely linked to its primary complication, myocardial infarction, ranking as the leading global cause of death \cite{WHO2020}. Additionally, we analyze atrial fibrillation (AF) (ICD$10$ code I$48$), a condition associated with heart failure \cite{Lip2007}, and diabetes mellitus (DM) (ICD$10$ codes E$10$-E$14$), which is associated with an increased risk for heart failure \cite{Cavender2015}. Because cardiovascular diseases can go undiagnosed for several years before they are recorded upon a severe cardiac event, we summarise both past and future diagnoses to address the challenge of accurately determining the disease onset. As all diseases are sparsely represented in the dataset ($7.22\,\%$ for CAD, $2.98\,\%$ for AF, and $4.46\,\%$ for DM), the fine-tune training sets were balanced using all positive subjects and a static set of randomly chosen negative subjects. 
\textcolor{black}{As a result, the balanced training sets comprise 4,030 subjects (2,015 positives) for CAD, 1,698 subjects (849 positives) for AF, and 2,498 subjects (1,249 positives) for DM.}
The validation and test sets were left untouched \textcolor{black}{to represent the disease prevalence in the population}.

To evaluate the potential of our method 
\textcolor{black}{for localised tasks,}
such as the prediction of cardiac imaging phenotypes, we 
\textcolor{black}{use a subset of} 
the \textcolor{black}{full} dataset to extract $61$ phenotypes from short-axis CMR images 
\textcolor{black}{following}
\cite{Bai2020}. 
\textcolor{black}{To this end, we employ stratified splits of the full dataset, resulting in 13,262 training, 2,835 validation, and 2,829 testing pairs, which preserve the distribution of the diseases in the population.}
\textcolor{black}{The cardiac phenotypes extracted from CMR images include}
the left ventricular (LV) end-diastolic volume (LVEDV) [mL], LV end-systolic volume (LVESV) [mL], LV stroke volume (LVSV) [mL], LV ejection fraction (LVEF) [\%], LV cardiac output (LVCO) [L/min], and LV mass (LVM) [g]. Furthermore, the right ventricular (RV) end-diastolic volume (RVEDV) [mL], RV end-systolic volume (RVESV) [mL], RV stroke volume (RVSV) [mL], and RV ejection fraction (RVEF) [\%] are extracted. Lastly, we determine the wall thickness (WT) [mm], circumferential strain (Ecc) [mm], and radial strain (Err) [mm], respectively, of all $17$ myocardial segments described by the American Heart Association (AHA) \cite{Cerqueira2002}.

\subsection{Experimental setup}
\label{sec:experimental_setup}
\subsubsection{Network architectures}
\label{sec:network_archs}
For the signal encoder $f_s(\cdot)$, we adapt a Vision Transformer (ViT) \cite{Dosovitskiy2020} with $3$ layers and $6$ heads that creates embeddings of size $384$, totalling $5\,$M model parameters. 
We reshape the $12$-lead ECG into a flattened sequence of one-dimensional patches with size $(1,100)$. 
For the image encoder, we use a ResNet50 \cite{He2016} that generates embeddings of size $2048$, totalling $24\,$M model parameters. 
All image encoders are pre-trained on the UK Biobank data using SimCLR \cite{Chen2020}, which is most performant for self-supervised unimodal pre-training of CMR images as shown by \cite{Hager2023}. 
Our signal and image projectors are two-layer perceptrons with one hidden layer of size $512$, which separately project the embeddings to a shared latent space of $128$ dimensions.

\subsubsection{Training strategy}
We first pre-train our unimodal signal encoder $f_s(\cdot)$ 
\textcolor{black}{on the $28,030$ training samples of our full dataset}
using the MAE \cite{He2022}. We use AdamW \cite{Loshchilov2017} with a batch size of $128$ and a cosine annealing scheduler over $400$ epochs with a $10\,\%$ warmup. We set a base learning rate of $10^{-5}$, a weight decay of $0.15$, and a masking ratio $\rho$ of $0.8$. We introduce an unstructured masking strategy for time series data similar to \cite{Huang2022}.

After pre-training solely on ECG data, we leverage multimodal contrastive learning to pre-train our signal encoder $f_s(\cdot)$ with information from CMR imaging. 
We use AdamW \cite{Loshchilov2017} with a weight decay of $10^{-4}$, a batch size of $256$ and a cosine annealing scheduler over $300$ epochs with a $10\,\%$ warmup. 
To find the optimal hyperparameters, we sweep over $\tau$ ($0.1$, $0.2$, $0.3$), $\lambda$ (0.25, 0.5, 0.75), and the learning rate ($3\cdot10^{-5}$, $10^{-4}$, $3\cdot10^{-4}$, $10^{-3}$).

Finally, we fine-tune our unimodal signal encoder $f_s(\cdot)$ to predict the subject-specific disease risk and cardiac imaging phenotypes solely from ECG data. 
To this end, we utilise the balanced training sets for disease prediction and the $13,262$ training samples for cardiac phenotype prediction, as detailed in Section \ref{sec:prediction_targets}. 
We use AdamW \cite{Loshchilov2017} and a cosine annealing scheduler over $400$ epochs with a $5\,\%$ warmup. We find optimal hyperparameters using a grid search, sweeping over the batch size ($8$, $16$, $32$), the layer decay ($0.5$, $0.75$), the drop path rate ($0.0$, $0.1$, $0.2$), the label smoothing rate ($0.0$, $0.1$, $0.2$), the weight decay ($0.0$, $0.1$, $0.2$), and the learning rate ($10^{-6}$, $3\cdot10^{-6}$, $10^{-5}$, $3\cdot10^{-5}$). We replace the global average pooling of $f_s(\cdot)$ used during pre-training with attention pooling as described in \cite{Radford2021}. We leave all model weights trainable during fine-tuning, as this typically yields better results compared to linear probing and thus is most common in practice. Fine-tuning is stopped using an early stopping callback on the validation metrics described in Section \ref{sec:metrics}.

We augment $12$-lead ECG using random cropping (scale=$0.5$ only during pre-training), Gaussian noise (sigma=$0.25$ during pre-training, $0.2$ during fine-tuning), amplitude rescaling (sigma=$0.5$ during pre-training and fine-tuning), and Fourier transform surrogates ($\text{phase\_noise\_magnitude}$=$0.1$ during pre-training, $0.075$ during fine-tuning). The CMR images are augmented using horizontal flips (probability=$0.5$), rotations (probability=$0.5$, degrees=$45$), colour jitter (brightness=$0.5$, contrast=$0.5$, saturation=$0.25$), and random resized cropping (size=$210$, scale=($0.6, 1$)). 
Data augmentation is applied only to the training samples, both during pre-training and fine-tuning.
We optimise all network parameters on one NVIDIA A40 GPU.

\subsection{Evaluation}
\subsubsection{Baseline methods}
We compare our approach with several fully supervised baselines, using either ECG, CMR imaging, or tabular data. 
\textcolor{black}{For the ECG baseline, we randomly initialise the ViT described in \ref{sec:network_archs} and train it fully supervised from scratch.}
The CMR imaging baseline is a randomly initialised ResNet50 \cite{He2016} that is trained fully supervised. 
For the tabular baseline, we use demographic data, namely age and sex, together with physiological data, namely weight and height, to predict the endpoints.
We use a XGBoost \cite{Chen2016} based tabular encoder and sweep over the number of boosting rounds ($50$, $75$, $100$, $125$, $150$), the maximum tree depth ($3$, $4$, $5$, $6$, $7$, $8$), the learning rate ($0.1$, $0.2$, $0.3$), and the subsample ratio ($0.5$, $0.75$, $1.0$) to find the best hyperparameter setting. We optimise a cross-entropy and MSE loss for the classification and regression tasks, respectively.

To evaluate the effectiveness of the MAE \cite{He2022} for learning meaningful ECG representations, we compare it to a fully supervised model and multiple unimodally pre-trained contrastive models. Therefore, we predict the subject-specific risk of cardiovascular diseases from ECG data using the fully supervised ViT \cite{Dosovitskiy2020} described in Section \ref{sec:network_archs} and the following contrastive approaches. 
SimCLR \cite{Chen2020} with a $\tau$ of $0.1$, learning rate of $3\cdot10^{-3}$, and weight decay of $10^{-4}$. 
BYOL \cite{Grill2020} with a $\tau$ of $0.9996$, learning rate of $10^{-5}$, and weight decay of $10^{-4}$. 
CLOCS \cite{Kiyasseh2021} with a $\tau$ of $0.1$, learning rate of $3\cdot10^{-4}$, and weight decay of $10^{-4}$. 
BarlowTwins \cite{Zbontar2021} with a learning rate of $3\cdot10^{-5}$ and weight decay of $1.5\cdot10^{-6}$.

\subsubsection{Metrics}
\label{sec:metrics}
We evaluate the methods' classification performance by calculating the 
area under the receiver operating characteristic curve (ROC AUC) as the UK Biobank study is severely unbalanced regarding the investigated cardiovascular diseases. The performance on the regression tasks is assessed using Pearson's correlation coefficient 
\textcolor{black}{($r$) and the coefficient of determination ($R^2$).} 
All experiments are reported as mean and standard deviation across five seeds set during fine-tuning.

\section{Results and discussion}
\subsection{Multimodal pre-training allows for CMR-level disease prediction solely from ECG}
\begin{table}[t!]
\caption{Comparison of different diagnostic modalities for the risk prediction of cardiovascular diseases, i.e., coronary artery disease (CAD), atrial fibrillation (AF), and diabetis mellitus (DM). Best scores (ROC AUC $[\%]$) are in \textbf{bold} font, second best \underline{underlined}.}
\label{tab:all_examinations}
\centering
\setlength{\tabcolsep}{1.5em}
\begin{tabular}{lccc}
	\toprule
	\multirow{2}{*}{\textbf{Modality}} & \multicolumn{3}{c}{\textbf{Cardiovascular disease}} \\
	\cmidrule(lr){2-4}
	{} & \textbf{CAD} & \textbf{AF} & \textbf{DM} \\
	\midrule
	ECG & 66.90$\pm$0.47 & 69.76$\pm$2.13 & 67.35$\pm$0.76 \\
	Tabular* & \underline{71.97$\pm$0.14} & 72.22$\pm$0.00 & 72.53$\pm$0.15\\
	CMR & 71.88$\pm$0.64 & \underline{73.11$\pm$0.64} & \underline{74.24$\pm$0.50} \\
	\midrule
	ECG (ours) & \textbf{73.00$\pm$0.30} & \textbf{74.11$\pm$0.53} & \textbf{75.56$\pm$0.16} \\
	\bottomrule
    \multicolumn{4}{l}{\scriptsize{*age, sex, height, and weight.}} \\
\end{tabular}
\end{table}

To evaluate the utility of our multimodal solution in a clinical setting, we compare it to baseline methods that use different modalities, including ECG, CMR imaging, and tabular data, for the subject-specific risk prediction of cardiovascular diseases. Each baseline is trained fully-supervised on data of the respective modality and subsequently tested with data of the same modality. Table \ref{tab:all_examinations} presents the performance of all models in predicting the risk of CAD, AF, and DM. Analyzing the results of the tabular baseline, we can see that cardiovascular diseases are highly correlated with the \textcolor{black}{investigated} demographic and physiological factors, \textcolor{black}{namely} age, sex, height, and weight. Furthermore, the experiments show that CMR imaging data is beneficial for predicting the risk of all diseases except CAD, that can be diagnosed as accurately with tabular data. Our proposed approach, that pre-trains multimodally but predicts the risk of diseases unimodally from ECG, consistently outperforms all the baselines, including the CMR imaging model.

These results indicate that leveraging large amounts of multimodal biobank data during pre-training eliminates the dependency on expensive and limited CMR imaging data during inference in clinical practice. In fact, our approach enables a subject-specific, holistic cardiac screening solely from cost-effective and widely available ECG.

\subsection{Multimodal pre-training with masked data modelling outperforms other pre-training strategies}
\label{sec:ecg_interpretation}
\begin{table*}[ht!]
\caption{Comparison of our approach against supervised and state-of-the-art self-supervised baseline models. Best scores (ROC AUC [$\%$]) are in \textbf{bold} font, second best \underline{underlined}. Our multimodally pre-trained model outperforms all baseline models on every task.}
\label{tab:ecg_um_comparison}
\centering
\setlength{\tabcolsep}{3.575em}
\begin{tabular}{llccc}
	\toprule
	\multirow{2}{*}{\textbf{Modality}} & \multirow{2}{*}{\textbf{Model}} & \multicolumn{3}{c}{\textbf{Cardiovascular disease}} \\
	\cmidrule(lr){3-5}
	{} & {} & \textbf{CAD} & \textbf{AF} & \textbf{DM} \\
	\midrule
	\multirow{8}{*}{ECG} & Supervised ViT & 66.90$\pm$0.47 & 69.76$\pm$2.13 & 67.35$\pm$0.76 \\
	\cmidrule{2-5}
	{} & SimCLR & 66.03$\pm$0.99 & 67.95$\pm$0.59 & 68.73$\pm$0.47 \\
	{} & CLOCS & 67.30$\pm$0.46 & 67.33$\pm$0.32 & 68.68$\pm$0.37 \\
	{} & BYOL & 67.59$\pm$0.28 & 69.76$\pm$0.52 & 67.70$\pm$0.26 \\
	{} & BarlowTwins & 68.11$\pm$0.54 & 67.42$\pm$0.72 & 69.58$\pm$0.02 \\
	{} & MAE & \underline{71.71$\pm$0.36} & \underline{72.32$\pm$0.73} & \underline{73.15$\pm$0.39} \\
	\cmidrule{2-5}
	{} & MMCL \textbf{w/} MDM (ours) & \textbf{73.00$\pm$0.30} & \textbf{74.11$\pm$0.53} & \textbf{75.56$\pm$0.16} \\
	\bottomrule
\end{tabular}
\end{table*}

\begin{table*}[ht!]
\caption{Ablation study on pre-training strategies for the risk prediction of cardiovascular diseases from ECG. Best scores (ROC AUC [$\%$]) are in \textbf{bold} font, second best \underline{underlined}. Multimodal contrastive learning with masked data modelling consistently outperforms all other methods.}
\label{tab:ecg_ablation}
\centering
\setlength{\tabcolsep}{2.125em}
\begin{tabular}{llccccc}
	\toprule
	\multirow{2}{*}{\textbf{Modality}} & \multirow{2}{*}{\textbf{Model}} & \multicolumn{2}{c}{\textbf{Pre-training}} & \multicolumn{3}{c}{\textbf{Cardiovascular disease}} \\
	\cmidrule(lr){3-4} \cmidrule(lr){5-7}
	{} & {} & \textbf{MDM} & \textbf{MMCL} & \textbf{CAD} & \textbf{AF} & \textbf{DM} \\
	\midrule
	\multirow{4}{*}{ECG} & Supervised ViT & \ding{55} & \ding{55} & 66.90$\pm$0.47 & 69.76$\pm$2.13 & 67.35$\pm$0.76 \\
	{} & MDM  & \ding{51} & \ding{55} & \underline{71.71$\pm$0.36} & 72.32$\pm$0.73 & 73.15$\pm$0.39 \\
	{} & MMCL & \ding{55} & \ding{51} & 70.36$\pm$0.51 & \underline{72.41$\pm$0.52} & \underline{73.40$\pm$0.48} \\
	{} & MMCL \textbf{w/} MDM (ours) & \ding{51} & \ding{51} & \textbf{73.00$\pm$0.30} & \textbf{74.11$\pm$0.53} & \textbf{75.56$\pm$0.16} \\
	\bottomrule
\end{tabular}
\end{table*}

\begin{table*}[ht!]
\caption{\textcolor{black}{Ablation study on established image encoders for visual feature extraction during multimodal pre-training. Best scores (ROC AUC [$\%$]) are in \textbf{bold} font. Convolutional networks are most effective in extracting general CMR imaging features that transfer well to ECG representations.}}
\label{tab:img_backbone_ablation}
\centering
\setlength{\tabcolsep}{2.675em}
{\color{black}\begin{tabular}{llcccc}
	\toprule
	\multirow{2}{*}{\textbf{Modality}} & \multirow{2}{*}{\textbf{Model}} & \multirow{2}{*}{\textbf{Image encoder}}& \multicolumn{3}{c}{\textbf{Cardiovascular disease}} \\
	\cmidrule(lr){4-6}
	{} & {} & {} & \textbf{CAD} & \textbf{AF} & \textbf{DM} \\
    \midrule
	\multirow{2}{*}{ECG} & MMCL \textbf{w/} MDM & ViT-B/16 & 71.97$\pm$0.31 & 72.33$\pm$1.15 & 75.51$\pm$0.52 \\
	{} & MMCL \textbf{w/} MDM & ResNet50 & \textbf{73.00$\pm$0.30} & \textbf{74.11$\pm$0.53} & \textbf{75.56$\pm$0.16} \\
	\bottomrule
\end{tabular}}
\end{table*}

\begin{table*}[t!]
\caption{Ablation study on pre-training strategies for predicting cardiac imaging phenotypes solely from ECG data. 
\textcolor{black}{Pearson's correlation coefficient ($r$) and the coefficient of determination ($R^2$) are reported as mean across all phenotypes.}
Best scores are in \textbf{bold} font, second best \underline{underlined}. Multimodal contrastive learning with masked data modelling consistently outperforms all other methods. The CMR imaging model is provided as reference.}
\label{tab:ecg_phenotype_ablation}
\centering
\setlength{\tabcolsep}{0.875em}
\begin{tabular}{llcccccccccccc}
	\toprule
	\multirow{3}{*}{\textbf{Modality}} & \multirow{3}{*}{\textbf{Model}} & \multicolumn{2}{c}{\textbf{Pre-training}} & \multicolumn{10}{c}{\textbf{Cardiac imaging phenotype}} \\
	\cmidrule(lr){3-4} \cmidrule(lr){5-14}
	{} & {} & \multirow{2}{*}{\textbf{MDM}} & \multirow{2}{*}{\textbf{MMCL}} & \multicolumn{2}{c}{\textbf{LV}} & \multicolumn{2}{c}{\textbf{RV}} & \multicolumn{2}{c}{\textbf{WT}} & \multicolumn{2}{c}{\textbf{Ecc}} & \multicolumn{2}{c}{\textbf{Err}} \\
	{} & {} & {} & {} & \footnotesize{$r$} & \textcolor{black}{\footnotesize{$R^2$}} & \footnotesize{$r$} & \textcolor{black}{\footnotesize{$R^2$}} & \footnotesize{$r$} & \textcolor{black}{\footnotesize{$R^2$}} & \footnotesize{$r$} & \textcolor{black}{\footnotesize{$R^2$}} & \footnotesize{$r$} & \textcolor{black}{\footnotesize{$R^2$}} \\
	\midrule
	\multirow{4}{*}{ECG} & Supervised ViT & \ding{55} & \ding{55} & 0.56 & \textcolor{black}{0.312} & 0.58 & \textcolor{black}{0.337} & 0.54 & \textcolor{black}{0.289} & 0.29 & \textcolor{black}{0.083} & 0.30 & \textcolor{black}{0.087} \\
	{} & MDM &  \ding{51} & \ding{55} & \underline{0.62} & \textcolor{black}{\underline{0.390}} & \underline{0.64} & \textcolor{black}{\underline{0.398}} & \underline{0.60} & \textcolor{black}{\underline{0.356}} & \underline{0.36} & \textcolor{black}{\underline{0.117}} & \underline{0.36} & \textcolor{black}{\underline{0.120}} \\
	{} & MMCL &  \ding{55} & \ding{51} & 0.58 & \textcolor{black}{0.335} & 0.60 & \textcolor{black}{0.351} & 0.57 & \textcolor{black}{0.325} & 0.32 & \textcolor{black}{0.109} & 0.33 & \textcolor{black}{0.107} \\
	{} & MMCL \textbf{w/} MDM (ours) &  \ding{51} & \ding{51} & \textbf{0.63} & \textcolor{black}{\textbf{0.407}} & \textbf{0.65} & \textcolor{black}{\textbf{0.421}} & \textbf{0.61} & \textcolor{black}{\textbf{0.381}} & \textbf{0.37} & \textcolor{black}{\textbf{0.142}} & \textbf{0.38} & \textcolor{black}{\textbf{0.139}} \\
	\midrule
	CMR$_\text{ref}$ & Supervised ResNet & \ding{55} & \ding{55} & 0.84 & \textcolor{black}{0.702} & 0.85 & \textcolor{black}{0.715} & 0.78 & \textcolor{black}{0.608} & 0.64 & \textcolor{black}{0.404} & 0.66 & \textcolor{black}{0.438} \\
	\bottomrule
\end{tabular}
\end{table*}

A comparison of our multimodal approach against supervised and multiple state-of-the-art self-supervised pre-training strategies can be found in Table \ref{tab:ecg_um_comparison}. We can observe that masked data modelling (MDM) with the MAE \cite{He2022} is the most effective unimodal pre-training strategy, noticeably outperforming all other unimodally pre-trained models across all diseases. Employing heavy masking ratios, we find that \textcolor{black}{MDM} is able to extract ECG representations with high information density that generalise well for detecting various cardiovascular diseases. 
\textcolor{black}{Despite masking $80\,$\% of the input signal, these dense ECG representations allow for a perfect reconstruction of the masked patches, indicated by a normalised MSE of $0.043$ across all leads and patches. This demonstrates the successful removal of redundancy inherent to standard 12-lead ECG.}
Our experiments show that contrastive techniques are less suitable for learning such meaningful ECG representations. We hypothesise that the augmentations described in Section \ref{sec:network_archs} are not meeting the requirement of heavy data augmentations, which are crucial for contrastive methods. Even contrastive implementations that use augmentations tailored to $12$-lead ECG data, such as CLOCS \cite{Kiyasseh2021}, are outperformed by \textcolor{black}{MDM}. These findings highlight our decision to integrate MDM into our proposed multimodal approach. Eventually, we can observe that multimodal contrastive learning (MMCL) combined with MDM, or short MMCL \textbf{w/} MDM, consistently outperforms all other methods.

To evaluate the strength of our approach, MMCL \textbf{w/} MDM, we perform an ablation study comparing different pre-training strategies for the risk prediction of diseases from ECG. We compare our approach against a supervised ViT, MDM, and MMCL. The results are presented in Table \ref{tab:ecg_ablation}. Our experiments show that MMCL generally improves performance compared to the unimodal baselines, resulting in a boost in prediction scores across all diseases except for CAD, where it is inferior to MDM. Further, we can observe that including MDM into the pre-training substantially improves upon MMCL, with MMCL \textbf{w/} MDM outperforming all other pre-training strategies on all three diseases. These findings suggest that the prior extraction of ECG representations with high information density facilitates the subsequent information transfer from CMR imaging.
\textcolor{black}{Furthermore, an ablation study comparing two established image encoders reveals that information from CMR imaging is best extracted with convolutional networks as shown in Table \ref{tab:img_backbone_ablation}. This indicates that the spatial inductive bias of convolutional architectures is crucial for our approach, especially because the number of paired ECG and CMR imaging data is limited.}

\subsection{Cardiac phenotype prediction from ECG benefits from multimodal pre-training}

\begin{figure*}[ht!]
    \centering
    \includegraphics[width=0.99\textwidth]{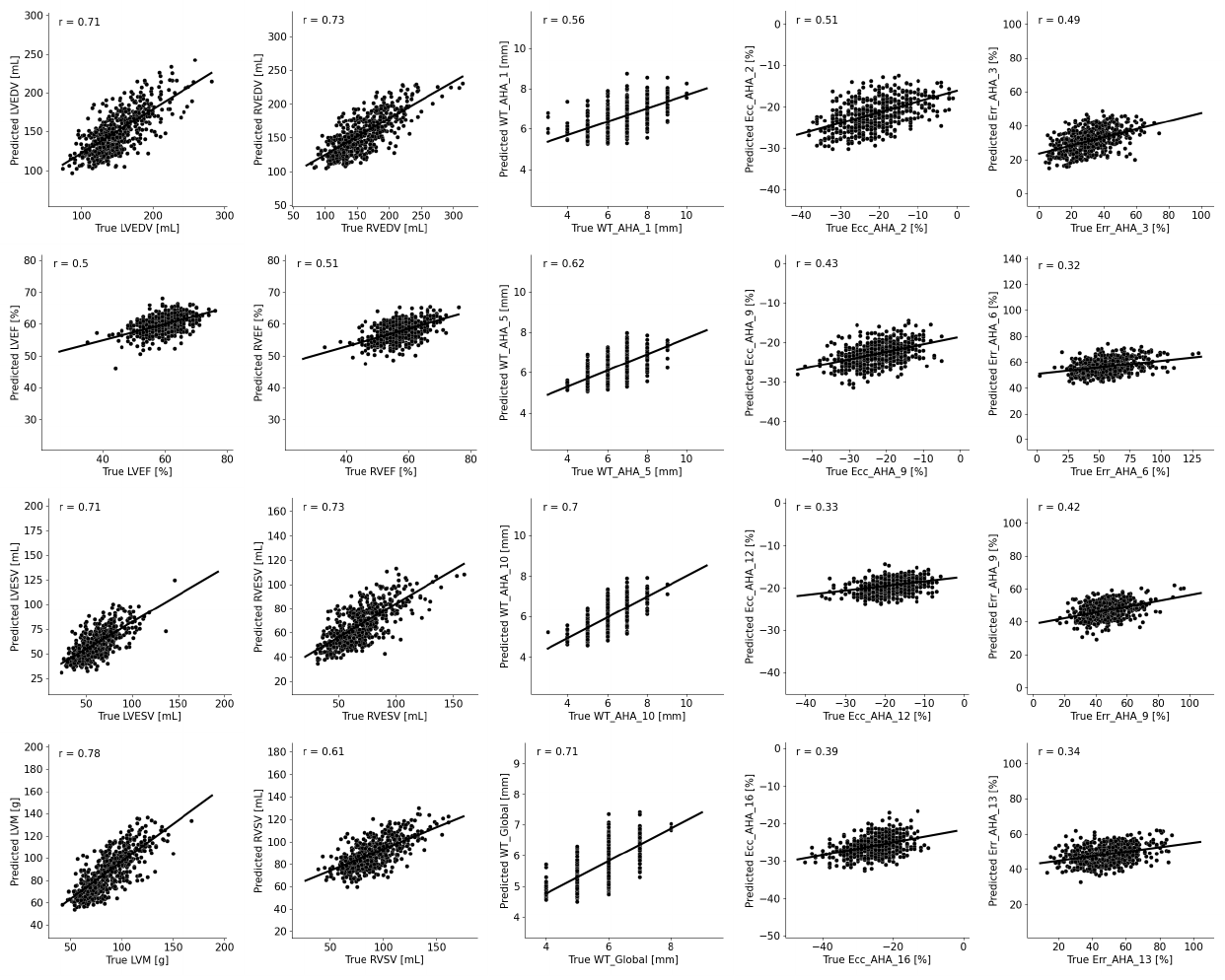}
    \caption{We use our multimodally pre-trained model to predict 61 cardiac imaging phenotypes solely from ECG data. The graphs show 20 imaging phenotypes of 500 subjects, as well as the linear regression line for the whole test set population. Pearson's correlation coefficient (r) is reported.}
    \label{fig:regression}
\end{figure*}

\begin{table*}[t!]
\caption{\textcolor{black}{Comparison of our approach against the state-of-the-art multimodal baseline. Best scores ($R^2$) are in \textbf{bold} font, second best \underline{underlined}. The Cross-Modal AE results are sourced from the original paper. Our multimodally pre-trained model outperforms the baseline model on every task.}}
\label{tab:ecg_phenotype_comparison}
\centering
\setlength{\tabcolsep}{0.71em}
{\color{black}\begin{tabular}{llccccccccccc}
	\toprule
	\multirow{2}{*}{\textbf{Modality}} & \multirow{2}{*}{\textbf{Model}} & \multicolumn{2}{c}{\textbf{Pre-training}} & \multicolumn{8}{c}{\textbf{Cardiac imaging phenotype}} \\
	\cmidrule(lr){3-4} \cmidrule(lr){5-13}
	{} & {} & \textbf{MDM} & \textbf{MMCL} & \textbf{LVEDV} & \textbf{LVESV} & \textbf{LVSV} & \textbf{LVEF} & \textbf{LVM} & \textbf{RVEDV} & \textbf{RVESV} & \textbf{RVSV} & \textbf{RVEF} \\
	\midrule
	\multirow{5}{*}{ECG} & Supervised ViT & \ding{55} & \ding{55} & 0.394 & 0.386 & 0.284 & 0.171 & 0.459 & 0.432 & 0.440 & 0.304 & 0.181 \\
	{} & MDM & \ding{51} & \ding{55} & \underline{0.478} & \underline{0.475} & \underline{0.349} & \underline{0.233} & \underline{0.565} & \underline{0.506} & \underline{0.508} & \underline{0.357} & \underline{0.230} \\
	{} & MMCL & \ding{55} & \ding{51} & 0.416 & 0.412 & 0.298 & 0.180 & 0.503 & 0.442 & 0.450 & 0.309 & 0.194 \\
	{} & MMCL \textbf{w/} MDM (ours) & \ding{51} & \ding{51} & \textbf{0.498} & \textbf{0.497} & \textbf{0.360} & \textbf{0.245} & \textbf{0.597} & \textbf{0.527} & \textbf{0.534} & \textbf{0.375} & \textbf{0.248} \\
	\cmidrule{2-13}
	{} & Cross-Modal AE* & $-$$^\dagger$ & \ding{51} & 0.451 & 0.380 & 0.316 & 0.103 & 0.536 & 0.490 & 0.445 & 0.320 & 0.129 \\
	\bottomrule
    \multicolumn{13}{l}{\scriptsize{* Pre-trained on $27,160$ ECG-CMR pairs. Fine-tuning performed on $3,613$ training and $527$ validation samples, with evaluation conducted on $528$ testing samples.}} \\
    \multicolumn{13}{l}{\scriptsize{$^\dagger$ In addition to the contrastive loss, the total loss used during multimodal pre-training also includes a reconstruction loss.}}
\end{tabular}}
\end{table*}

\begin{figure*}[ht!]
    \centering
    \includegraphics[width=0.965\textwidth]{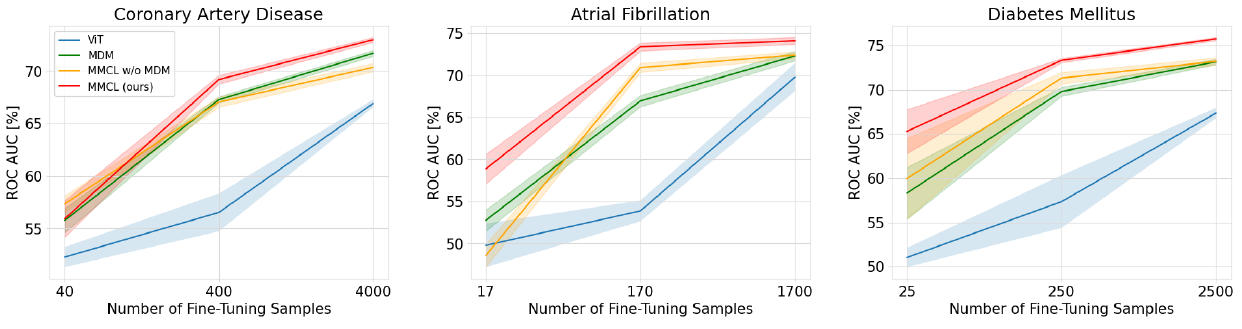}
    \caption{Performance of our multimodal approach with different number of fine-tune training samples. We compare our solution to supervised and self-supervised baseline models. Shaded regions indicate 95\,\% confidence intervals. Multimodal contrastive learning with masked data modelling generally outperforms all other models at all data quantities and is well suited for predicting risks of rare diseases.}
    \label{fig:low_data}
\end{figure*}

\begin{figure*}[ht!]
    \centering
    \includegraphics[width=0.75\textwidth]{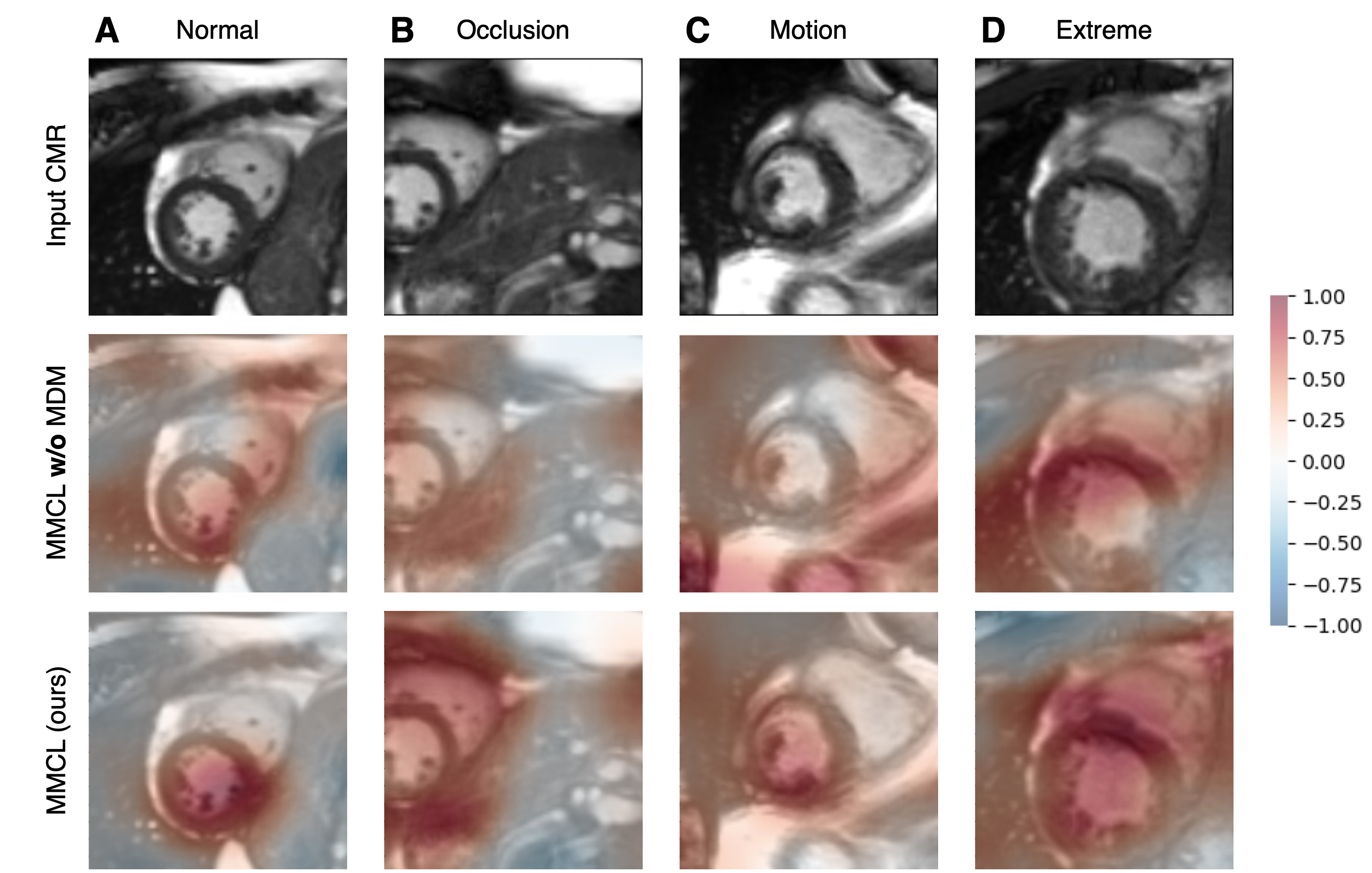}
    \caption{Qualitative evaluation of the information transfer from CMR imaging to ECG using our patch-based interpretability module. Cosine similarities between the ECG representation and representations of local CMR image regions produce a similarity grid that is superimposed onto the CMR image as heatmap. Our approach elicits high similarity between the ECG and CMR image regions of interest and achieves robust information transfer from CMR images even in the cases of occlusion, motion, and samples showing extreme morphological properties.}
    \label{fig:alignment}
\end{figure*}

We further investigate the extent to which morphological information is encoded in ECG. 
Specifically, we predict distinct cardiac phenotypes that are derived from short-axis CMR images, as summarised in Table \ref{tab:ecg_phenotype_ablation}. 
The experiments show that while MMCL enhances the prediction of cardiac imaging phenotypes, it is less effective than MDM.
These findings indicate that morphological features are inherently present in $12$-lead ECG and can be better accessed by eliminating redundancy across signal channels and time, stressing the value of masking as a robust time series augmentation technique. 
Notably, MMCL \textbf{w/} MDM proves most effective in predicting cardiac imaging phenotypes, highlighting the importance of MDM in facilitating information transfer from CMR imaging to ECG during multimodal pre-training. In Figure \ref{fig:regression}, we demonstrate the ability of our multimodally pre-trained model to predict cardiac phenotypes solely from ECG.

Additionally, we compare our approach against the multimodally pre-trained model proposed by \cite{Radhakrishnan2023}, referred to here as Cross-Modal AE. 
To this end, we evaluate the model performance across nine cardiac imaging phenotypes, including five LV and four RV phenotypes, as detailed in Table \ref{tab:ecg_phenotype_comparison}.
The results reveal that our approach consistently outperforms the baseline across all phenotypes, stressing the effectiveness of our multimodal pre-training paradigm.
Overall, these findings demonstrate that our method generates \textit{general-purpose} cardiac representations from ECG, which are not only suitable for risk prediction of cardiac diseases but also translate well to predicting important cardiac phenotypes. 

\subsection{Multimodal pre-training is particularly powerful in low-data regimes}
Investigating medical conditions characterised by low prevalence requires the model performance to be robust even when trained with few positive samples. Hence, to evaluate the effectiveness of our multimodal method under low-data conditions, we conduct experiments by subsampling the fine-tuning training set to $10\,\%$ and $1\,\%$ of its original size, while ensuring that each subset is fully contained within the corresponding superset. Consequently, the balanced training set sizes for CAD, AF, and DM were $400$ ($10\,\%$) and $40$ ($1\,\%$), $170$ ($10\,\%$) and $17$ ($1\,\%$), and $250$ ($10\,\%$) and $25$ ($1\,\%$), respectively. The validation and test sets remain unchanged from the full data regime. Figure \ref{fig:low_data} illustrates the performance of our approach as well as the baselines under low-data conditions.

Our results indicate that in low-data regimes, MMCL \textbf{w/} MDM consistently outperforms the baselines across all diseases, except in the detection of CAD with $1\,\%$ fine-tuning samples. This finding highlights the enhanced ECG representations generated by our multimodal approach, which requires fewer fine-tuning samples to achieve competitive performance especially when investigating rare diseases.

\begin{table*}[ht!]
\caption{Ablation study on pre-training strategies for the risk prediction of cardiovascular diseases from CMR images. Best scores (ROC AUC $[\%]$) are in \textbf{bold} font, second best \underline{underlined}. Unimodal CMR image analysis benefits from multimodal pre-training.}
\label{tab:cmr_ablation}
\centering
\setlength{\tabcolsep}{2.15em}
\begin{tabular}{llccccc}
	\toprule
	\multirow{2}{*}{\textbf{Modality}} & \multirow{2}{*}{\textbf{Model}} & \multicolumn{2}{c}{\textbf{Pre-training}} & \multicolumn{3}{c}{\textbf{Cardiovascular disease}} \\
	\cmidrule(lr){3-4} \cmidrule(lr){5-7}
	{} & {} & \textbf{MDM} & \textbf{MMCL} & \textbf{CAD} & \textbf{AF} & \textbf{DM} \\
	\midrule
	\multirow{3}{*}{CMR} & Supervised ResNet & \ding{55} & \ding{55} & 71.88$\pm$0.64 & 73.11$\pm$0.64 & 74.24$\pm$0.50 \\
	{} & MMCL & \ding{55} & \ding{51} & \textbf{74.69$\pm$0.23} & \textbf{77.36$\pm$0.62} & \underline{77.37$\pm$0.42} \\
	{} & MMCL \textbf{w/} MDM (ours) & \ding{51} & \ding{51} & \underline{74.37$\pm$0.69} & \underline{77.13$\pm$0.42} & \textbf{77.82$\pm$0.26} \\
	\bottomrule
\end{tabular}
\end{table*}
\subsection{Patch-based interpretability module can visualise information transfer from CMR imaging to ECG}
To further explore why incorporating MDM into MMCL improves both the subject-specific risk prediction of diseases and the prediction of cardiac imaging phenotypes, we conduct a qualitative analysis of the information transfer from CMR imaging to ECG. 
To this end, we use our patch-based interpretability module 
\textcolor{black}{to visualise the information transfer, as depicted}
in Figure \ref{fig:alignment}. 
We find that leveraging MDM elicits higher correlations between the ECG and CMR image regions of interest, such as the left ventricle. These findings are also in line with existing medical literature, which links cardiovascular diseases to the left ventricle \cite{Davies2001}. Furthermore, our analysis indicates that MDM allows for a robust information transfer even in the cases of occlusion, motion, and samples showing extreme morphological properties. Taken together, when incorporating MDM into MMCL, we can extract meaningful ECG representations that include structural information from CMR imaging. This eventually leads to a performance boost in risk prediction as well as cardiac phenotype prediction.

\subsection{Multimodal pre-training also improves disease prediction from CMR imaging}
To evaluate the information transfer from ECG to CMR imaging, we perform additional experiments to predict the risk of cardiovascular diseases from CMR images. In Table \ref{tab:cmr_ablation}, we compare our proposed approach MMCL \textbf{w/} MDM to MMCL and a fully supervised ResNet50 \cite{He2016}. The results show that our multimodal solutions substantially boost risk prediction performance, demonstrating a noticeable transfer of information also from ECG to CMR imaging. Moreover, while MDM is essential for the information transfer from CMR imaging to ECG, we find that it is not required for multimodally learning rich CMR imaging representations.

\section{Conclusion}
In this work, we present a novel contrastive learning approach that can pre-train multimodally using $12$-lead ECG and CMR imaging data. During inference, our solution requires only ECG data to unimodally assess subject-specific risks of cardiovascular diseases and to determine distinct cardiac phenotypes. Specifically, we combine multimodal contrastive learning with masked data modelling to transfer information from CMR imaging to ECG. We demonstrate both quantitatively and qualitatively that an information transfer from CMR imaging unlocks the full diagnostic potential of ECG, enabling affordable care for patients with cardiovascular conditions.

In the end, we also acknowledge limitations of our work. While our approach can predict unimodally solely from ECG data, it relies on paired ECG and CMR imaging data during pre-training, which are rare outside of large biobanks. A further shortcoming of this study is that mainly healthy subjects are included, since the investigated diseases are low frequency in the general population. The UK Biobank study also mainly considers white subjects, with other ethnicities making up only $5\,\%$ of the total cohort. The behaviour of such frameworks with more balanced data should be investigated in future work. 
Furthermore, our results indicate that the addition of local alignment between ECG time points and CMR imaging frames may be worth investigating in future work.

Overall, this study has demonstrated that multimodal contrastive learning in combination with mask data modelling generates general-purpose ECG representations that can be applied to a broad range of clinical downstream applications. Thus, we present a simple yet effective strategy that combines the accessibility of ECG with the informative value of CMR imaging, enabling holistic cardiac screening solely by ECG.

\bibliographystyle{ieeetr}
\bibliography{tmi.bib}

\begin{thebibliography}{10}

\bibitem{Siontis2021}
K.~C. Siontis, P.~A. Noseworthy, Z.~I. Attia, and P.~A. Friedman, ``Artificial intelligence-enhanced electrocardiography in cardiovascular disease management,'' {\em Nature Reviews Cardiology}, vol.~18, no.~7, pp.~465--478, 2021.

\bibitem{Gulrajani1998}
R.~M. Gulrajani, ``The forward and inverse problems of electrocardiography,'' {\em IEEE Engineering in Medicine and Biology Magazine}, vol.~17, no.~5, pp.~84--101, 1998.

\bibitem{Schenone2016}
E.~Schenone, A.~Collin, and J.-F. Gerbeau, ``Numerical simulation of electrocardiograms for full cardiac cycles in healthy and pathological conditions,'' {\em International journal for numerical methods in biomedical engineering}, vol.~32, no.~5, p.~e02744, 2016.

\bibitem{Engelen1999}
D.~J. Engelen, A.~P. Gorgels, E.~C. Cheriex, {\em et~al.}, ``Value of the electrocardiogram in localizing the occlusion site in the left anterior descending coronary artery in acute anterior myocardial infarction,'' {\em Journal of the American College of Cardiology}, vol.~34, no.~2, pp.~389--395, 1999.

\bibitem{Xiong2021}
P.~Xiong, Y.~Xue, J.~Zhang, {\em et~al.}, ``Localization of myocardial infarction with multi-lead ecg based on densenet,'' {\em Computer Methods and Programs in Biomedicine}, vol.~203, p.~106024, 2021.

\bibitem{Pombo1971}
J.~F. Pombo, B.~L. Troy, and R.~O. Russell~JR, ``Left ventricular volumes and ejection fraction by echocardiography,'' {\em Circulation}, vol.~43, no.~4, pp.~480--490, 1971.

\bibitem{Avendi2016}
M.~R. Avendi, A.~Kheradvar, and H.~Jafarkhani, ``A combined deep-learning and deformable-model approach to fully automatic segmentation of the left ventricle in cardiac mri,'' {\em Medical image analysis}, vol.~30, pp.~108--119, 2016.

\bibitem{Duffy2022}
G.~Duffy, P.~P. Cheng, N.~Yuan, {\em et~al.}, ``High-throughput precision phenotyping of left ventricular hypertrophy with cardiovascular deep learning,'' {\em JAMA cardiology}, vol.~7, no.~4, pp.~386--395, 2022.

\bibitem{Lee2018}
D.~C. Lee, M.~Markl, E.~Dall’Armellina, {\em et~al.}, ``The growth and evolution of cardiovascular magnetic resonance: a 20-year history of the society for cardiovascular magnetic resonance ({SCMR}) annual scientific sessions,'' {\em Journal of Cardiovascular Magnetic Resonance}, vol.~20, no.~1, pp.~1--11, 2018.

\bibitem{Reynolds2021}
H.~R. Reynolds, A.~Maehara, R.~Y. Kwong, {\em et~al.}, ``Coronary optical coherence tomography and cardiac magnetic resonance imaging to determine underlying causes of myocardial infarction with nonobstructive coronary arteries in women,'' {\em Circulation}, vol.~143, no.~7, pp.~624--640, 2021.

\bibitem{Wang2024}
Y.-R. Wang, K.~Yang, Y.~Wen, {\em et~al.}, ``Screening and diagnosis of cardiovascular disease using artificial intelligence-enabled cardiac magnetic resonance imaging,'' {\em Nature Medicine}, pp.~1--10, 2024.

\bibitem{Oakes2009}
R.~S. Oakes, T.~J. Badger, E.~G. Kholmovski, {\em et~al.}, ``Detection and quantification of left atrial structural remodeling with delayed-enhancement magnetic resonance imaging in patients with atrial fibrillation,'' {\em Circulation}, vol.~119, no.~13, pp.~1758--1767, 2009.

\bibitem{Bertelsen2020}
L.~Bertelsen, S.~Z. Diederichsen, K.~J. Haugan, {\em et~al.}, ``Left atrial volume and function assessed by cardiac magnetic resonance imaging are markers of subclinical atrial fibrillation as detected by continuous monitoring,'' {\em EP Europace}, vol.~22, no.~5, pp.~724--731, 2020.

\bibitem{Emerging2010}
E.~R.~F. Collaboration {\em et~al.}, ``Diabetes mellitus, fasting blood glucose concentration, and risk of vascular disease: a collaborative meta-analysis of 102 prospective studies,'' {\em The lancet}, vol.~375, no.~9733, pp.~2215--2222, 2010.

\bibitem{Sorensen2020}
M.~H. S{\o}rensen, A.~S. Bojer, D.~A. Broadbent, {\em et~al.}, ``Cardiac perfusion, structure, and function in type 2 diabetes mellitus with and without diabetic complications,'' {\em European Heart Journal-Cardiovascular Imaging}, vol.~21, no.~8, pp.~887--895, 2020.

\bibitem{Ng2021}
A.~C. Ng, V.~Delgado, B.~A. Borlaug, and J.~J. Bax, ``Diabesity: the combined burden of obesity and diabetes on heart disease and the role of imaging,'' {\em Nature Reviews Cardiology}, vol.~18, no.~4, pp.~291--304, 2021.

\bibitem{Von2017}
F.~von Knobelsdorff-Brenkenhoff, G.~Pilz, and J.~Schulz-Menger, ``Representation of cardiovascular magnetic resonance in the {AHA}/{ACC} guidelines,'' {\em Journal of Cardiovascular Magnetic Resonance}, vol.~19, no.~1, pp.~1--21, 2017.

\bibitem{Sudlow2015}
C.~Sudlow, J.~Gallacher, N.~Allen, {\em et~al.}, ``{UK} biobank: an open access resource for identifying the causes of a wide range of complex diseases of middle and old age,'' {\em PLoS Medicine}, vol.~12, no.~3, p.~e1001779, 2015.

\bibitem{Attia2019}
Z.~I. Attia, S.~Kapa, F.~Lopez-Jimenez, {\em et~al.}, ``Screening for cardiac contractile dysfunction using an artificial intelligence--enabled electrocardiogram,'' {\em Nature Medicine}, vol.~25, no.~1, pp.~70--74, 2019.

\bibitem{Raghunath2020}
S.~Raghunath, A.~E. Ulloa~Cerna, L.~Jing, {\em et~al.}, ``Prediction of mortality from 12-lead electrocardiogram voltage data using a deep neural network,'' {\em Nature Medicine}, vol.~26, no.~6, pp.~886--891, 2020.

\bibitem{Zhu2020}
H.~Zhu, C.~Cheng, H.~Yin, {\em et~al.}, ``Automatic multilabel electrocardiogram diagnosis of heart rhythm or conduction abnormalities with deep learning: a cohort study,'' {\em The Lancet Digital Health}, vol.~2, no.~7, pp.~e348--e357, 2020.

\bibitem{Sarkar2020}
P.~Sarkar and A.~Etemad, ``Self-supervised {ECG} representation learning for emotion recognition,'' {\em IEEE Transactions on Affective Computing}, vol.~13, no.~3, pp.~1541--1554, 2020.

\bibitem{Zhang2022}
H.~Zhang, W.~Liu, J.~Shi, {\em et~al.}, ``{MaeFE}: Masked autoencoders family of electrocardiogram for self-supervised pretraining and transfer learning,'' {\em IEEE Transactions on Instrumentation and Measurement}, vol.~72, pp.~1--15, 2022.

\bibitem{Kiyasseh2021}
D.~Kiyasseh, T.~Zhu, and D.~A. Clifton, ``Clocs: Contrastive learning of cardiac signals across space, time, and patients,'' in {\em International Conference on Machine Learning}, pp.~5606--5615, PMLR, 2021.

\bibitem{Khurshid2022}
S.~Khurshid, S.~Friedman, C.~Reeder, {\em et~al.}, ``Ecg-based deep learning and clinical risk factors to predict atrial fibrillation,'' {\em Circulation}, vol.~145, no.~2, pp.~122--133, 2022.

\bibitem{Yuan2023}
N.~Yuan, G.~Duffy, S.~S. Dhruva, {\em et~al.}, ``Deep learning of electrocardiograms in sinus rhythm from us veterans to predict atrial fibrillation,'' {\em JAMA cardiology}, vol.~8, no.~12, pp.~1131--1139, 2023.

\bibitem{Jahmunah2021}
V.~Jahmunah, E.~Y.~K. Ng, T.~R. San, and U.~R. Acharya, ``Automated detection of coronary artery disease, myocardial infarction and congestive heart failure using gaborcnn model with ecg signals,'' {\em Computers in biology and medicine}, vol.~134, p.~104457, 2021.

\bibitem{Awasthi2023}
S.~Awasthi, N.~Sachdeva, Y.~Gupta, {\em et~al.}, ``Identification and risk stratification of coronary disease by artificial intelligence-enabled ecg,'' {\em EClinicalMedicine}, vol.~65, 2023.

\bibitem{Khurshid2021}
S.~Khurshid, S.~Friedman, and J.~P.~a. Pirruccello, ``Deep learning to predict cardiac magnetic resonance--derived left ventricular mass and hypertrophy from 12-lead ecgs,'' {\em Circulation: Cardiovascular Imaging}, vol.~14, no.~6, p.~e012281, 2021.

\bibitem{Vaid2023}
A.~Vaid, J.~Jiang, A.~Sawant, {\em et~al.}, ``A foundational vision transformer improves diagnostic performance for electrocardiograms,'' {\em NPJ Digital Medicine}, vol.~6, no.~1, p.~108, 2023.

\bibitem{Chen2020Generative}
M.~Chen, A.~Radford, R.~Child, {\em et~al.}, ``Generative pretraining from pixels,'' in {\em International conference on machine learning}, pp.~1691--1703, PMLR, 2020.

\bibitem{Pathak2016}
D.~Pathak, P.~Krahenbuhl, J.~Donahue, {\em et~al.}, ``Context encoders: Feature learning by inpainting,'' in {\em IEEE/CVF Conference on Computer Vision and Pattern Recognition}, pp.~2536--2544, 2016.

\bibitem{Vincent2010}
P.~Vincent, H.~Larochelle, I.~Lajoie, {\em et~al.}, ``Stacked denoising autoencoders: Learning useful representations in a deep network with a local denoising criterion.,'' {\em Journal of machine learning research}, vol.~11, no.~12, 2010.

\bibitem{Devlin2018}
J.~Devlin, M.-W. Chang, K.~Lee, and K.~Toutanova, ``Bert: Pre-training of deep bidirectional transformers for language understanding,'' {\em North American Chapter of the Association for Computational Linguistics: Human Language Technologies}, pp.~4171--4186, 2019.

\bibitem{He2022}
K.~He, X.~Chen, S.~Xie, {\em et~al.}, ``Masked autoencoders are scalable vision learners,'' in {\em IEEE/CVF Conference on Computer Vision and Pattern Recognition}, pp.~16000--16009, 2022.

\bibitem{Feichtenhofer2022}
C.~Feichtenhofer, H.~Fan, Y.~Li, and K.~He, ``Masked autoencoders as spatiotemporal learners,'' in {\em Advances in Neural Information Processing Systems}, 2022.

\bibitem{Huang2022}
P.-Y. Huang, H.~Xu, J.~Li, {\em et~al.}, ``Masked autoencoders that listen,'' in {\em Advances in Neural Information Processing Systems}, 2022.

\bibitem{Chen2020}
T.~Chen, S.~Kornblith, M.~Norouzi, and G.~Hinton, ``A simple framework for contrastive learning of visual representations,'' in {\em International Conference on Machine Learning}, pp.~1597--1607, PMLR, 2020.

\bibitem{Assran2023}
M.~Assran, Q.~Duval, I.~Misra, {\em et~al.}, ``Self-supervised learning from images with a joint-embedding predictive architecture,'' in {\em IEEE/CVF Conference on Computer Vision and Pattern Recognition}, pp.~15619--15629, 2023.

\bibitem{Brown2020}
T.~Brown, B.~Mann, N.~Ryder, {\em et~al.}, ``Language models are few-shot learners,'' {\em Advances in neural information processing systems}, vol.~33, pp.~1877--1901, 2020.

\bibitem{Dosovitskiy2020}
A.~Dosovitskiy, L.~Beyer, A.~Kolesnikov, {\em et~al.}, ``An image is worth 16x16 words: Transformers for image recognition at scale,'' {\em International Conference on Learning Representations}, 2021.

\bibitem{Bao2022}
H.~Bao, L.~Dong, S.~Piao, {\em et~al.}, ``Beit: {BERT} pre-training of image transformers,'' in {\em International Conference on Learning Representations}, 2022.

\bibitem{Vaswani2017}
A.~Vaswani, N.~Shazeer, N.~Parmar, {\em et~al.}, ``Attention is all you need,'' {\em Advances in neural information processing systems}, vol.~30, 2017.

\bibitem{Beltagy2020}
I.~Beltagy, M.~E. Peters, and A.~Cohan, ``Longformer: The long-document transformer,'' {\em arXiv:2004.05150}, 2020.

\bibitem{Grill2020}
J.-B. Grill, F.~Strub, F.~Altch{\'e}, {\em et~al.}, ``Bootstrap your own latent-a new approach to self-supervised learning,'' {\em Advances in neural information processing systems}, vol.~33, pp.~21271--21284, 2020.

\bibitem{Zbontar2021}
J.~Zbontar, L.~Jing, I.~Misra, {\em et~al.}, ``Barlow twins: Self-supervised learning via redundancy reduction,'' in {\em International Conference on Machine Learning}, pp.~12310--12320, PMLR, 2021.

\bibitem{Hadsell2006}
R.~Hadsell, S.~Chopra, and Y.~LeCun, ``Dimensionality reduction by learning an invariant mapping,'' in {\em IEEE/CVF Conference on Computer Vision and Pattern Recognition}, vol.~2, pp.~1735--1742, IEEE, 2006.

\bibitem{Radford2021}
A.~Radford, J.~W. Kim, and C.~a. Hallacy, ``Learning transferable visual models from natural language supervision,'' in {\em International Conference on Machine Learning}, pp.~8748--8763, PMLR, 2021.

\bibitem{Zhang2022conVIRT}
Y.~Zhang, H.~Jiang, Y.~Miura, {\em et~al.}, ``Contrastive learning of medical visual representations from paired images and text,'' in {\em Machine Learning for Healthcare Conference}, pp.~2--25, PMLR, 2022.

\bibitem{Pielawski2020}
N.~Pielawski, E.~Wetzer, J.~{\"O}fverstedt, {\em et~al.}, ``Comir: Contrastive multimodal image representation for registration,'' {\em Advances in neural information processing systems}, vol.~33, pp.~18433--18444, 2020.

\bibitem{Hager2023}
P.~Hager, M.~J. Menten, and D.~Rueckert, ``Best of both worlds: Multimodal contrastive learning with tabular and imaging data,'' in {\em IEEE/CVF Conference on Computer Vision and Pattern Recognition}, pp.~23924--23935, 2023.

\bibitem{Xu2021}
H.~Xu, G.~Ghosh, P.-Y. Huang, {\em et~al.}, ``Videoclip: Contrastive pre-training for zero-shot video-text understanding,'' in {\em Conference on Empirical Methods in Natural Language Processing (EMNLP)}, 2021.

\bibitem{Zolfaghari2021}
M.~Zolfaghari, Y.~Zhu, P.~Gehler, and T.~Brox, ``Crossclr: Cross-modal contrastive learning for multi-modal video representations,'' in {\em IEEE/CVF International Conference on Computer Vision}, pp.~1450--1459, 2021.

\bibitem{Wang2021}
L.~Wang, P.~Luc, A.~Recasens, {\em et~al.}, ``Multimodal self-supervised learning of general audio representations,'' {\em arXiv preprint arXiv:2104.12807}, 2021.

\bibitem{Ma2021}
S.~Ma, Z.~Zeng, D.~McDuff, and Y.~Song, ``Active contrastive learning of audio-visual video representations,'' in {\em International Conference on Learning Representations}, 2021.

\bibitem{Radhakrishnan2023}
A.~Radhakrishnan, S.~F. Friedman, S.~Khurshid, {\em et~al.}, ``Cross-modal autoencoder framework learns holistic representations of cardiovascular state,'' {\em Nature Communications}, vol.~14, no.~1, p.~2436, 2023.

\bibitem{Zhang2022Exp}
Q.~Zhang, Y.~Wang, and Y.~Wang, ``How mask matters: Towards theoretical understandings of masked autoencoders,'' {\em Advances in Neural Information Processing Systems}, 2022.

\bibitem{Oord2018}
A.~v.~d. Oord, Y.~Li, and O.~Vinyals, ``Representation learning with contrastive predictive coding,'' {\em arXiv preprint arXiv:1807.03748}, 2018.

\bibitem{Huang2021}
S.-C. Huang, L.~Shen, M.~P. Lungren, and S.~Yeung, ``Gloria: A multimodal global-local representation learning framework for label-efficient medical image recognition,'' in {\em IEEE/CVF Conference on Computer Vision and Pattern Recognition}, pp.~3942--3951, 2021.

\bibitem{Boecking2022}
B.~Boecking, N.~Usuyama, S.~Bannur, {\em et~al.}, ``Making the most of text semantics to improve biomedical vision--language processing,'' in {\em Computer Vision--ECCV 2022}, pp.~1--21, Springer, 2022.

\bibitem{Zhang2010}
Z.-M. Zhang, S.~Chen, and Y.-Z. Liang, ``Baseline correction using adaptive iteratively reweighted penalized least squares,'' {\em Analyst}, vol.~135, no.~5, pp.~1138--1146, 2010.

\bibitem{Khan2008}
M.~G. Khan, {\em Rapid ECG Interpretation}.
\newblock Springer, 2008.

\bibitem{Raisi2021}
Z.~Raisi-Estabragh, N.~C. Harvey, S.~Neubauer, and S.~E. Petersen, ``Cardiovascular magnetic resonance imaging in the uk biobank: a major international health research resource,'' {\em European Heart Journal-Cardiovascular Imaging}, vol.~22, no.~3, pp.~251--258, 2021.

\bibitem{Tsao2015}
C.~W. Tsao, P.~N. Gona, C.~J. Salton, {\em et~al.}, ``Left ventricular structure and risk of cardiovascular events: a framingham heart study cardiac magnetic resonance study,'' {\em Journal of the American Heart Association}, vol.~4, no.~9, p.~e002188, 2015.

\bibitem{WHO2020}
WHO, ``Who methods and data sources for country-level causes of death 2000-2019,'' {\em World Health Organization}, 2020.

\bibitem{Lip2007}
G.~Y. Lip and H.-F. Tse, ``Management of atrial fibrillation,'' {\em The Lancet}, vol.~370, no.~9587, pp.~604--618, 2007.

\bibitem{Cavender2015}
M.~A. Cavender, P.~G. Steg, S.~C. Smith~Jr, {\em et~al.}, ``Impact of diabetes mellitus on hospitalization for heart failure, cardiovascular events, and death: outcomes at 4 years from the reduction of atherothrombosis for continued health (reach) registry,'' {\em Circulation}, vol.~132, no.~10, pp.~923--931, 2015.

\bibitem{Bai2020}
W.~Bai, H.~Suzuki, J.~Huang, {\em et~al.}, ``A population-based phenome-wide association study of cardiac and aortic structure and function,'' {\em Nature medicine}, vol.~26, no.~10, pp.~1654--1662, 2020.

\bibitem{Cerqueira2002}
A.~H. A. W.~G. on~Myocardial~Segmentation, R.~for Cardiac~Imaging:, M.~D. Cerqueira, N.~J. Weissman, V.~Dilsizian, A.~K. Jacobs, S.~Kaul, W.~K. Laskey, D.~J. Pennell, J.~A. Rumberger, T.~Ryan, and M.~S. Verani, ``Standardized myocardial segmentation and nomenclature for tomographic imaging of the heart: a statement for healthcare professionals from the cardiac imaging committee of the council on clinical cardiology of the american heart association,'' {\em Circulation}, vol.~105, no.~4, pp.~539--542, 2002.

\bibitem{He2016}
K.~He, X.~Zhang, S.~Ren, and J.~Sun, ``Deep residual learning for image recognition,'' in {\em IEEE/CVF Conference on Computer Vision and Pattern Recognition}, pp.~770--778, 2016.

\bibitem{Loshchilov2017}
I.~Loshchilov and F.~Hutter, ``Decoupled weight decay regularization,'' {\em International Conference on Learning Representations}, 2019.

\bibitem{Chen2016}
T.~Chen and C.~Guestrin, ``Xgboost: A scalable tree boosting system,'' in {\em Proceedings of the 22nd ACM SIGKDD International Conference on Knowledge Discovery and Data Mining}, pp.~785--794, 2016.

\bibitem{Davies2001}
M.~Davies, F.~Hobbs, R.~Davis, {\em et~al.}, ``Prevalence of left-ventricular systolic dysfunction and heart failure in the echocardiographic heart of england screening study: a population based study,'' {\em The Lancet}, vol.~358, no.~9280, pp.~439--444, 2001.

\end{thebibliography}

\end{document}